# Parts of Speech Tagging in NLP – Runtime Optimization with Quantum Formulation and ZX Calculus

ARIT KUMAR BISHWAS, Amity Institute of Information Technology, Noida, India
ASHISH MANI, Department of EEE, Amity University, Noida, India
VASILE PALADE, Research Centre for Data Science, Coventry University, Coventry, UK

This paper proposes an optimized formulation of the parts of speech tagging in Natural Language Processing with a quantum computing approach and further demonstrates the quantum gate-level runnable optimization with ZX-calculus, keeping the implementation target in the context of Noisy Intermediate Scale Quantum Systems (NISQ). Our quantum formulation exhibits quadratic speed up over the classical counterpart and further demonstrates the implementable optimization with the help of ZX calculus postulates.

## 1 INTRODUCTION

In the present time, the progress in developing quantum computers is very impressive. Many organizations are claiming their stacks in this space **[1][2][3][4]**. In today's world, the available quantum computers are at very early stages and not capable of handling complex quantum artificial intelligence/machine learning (qAI/qML) tasks **[5]**. But we still can harness their properties to run some of our quantum AI/ML algorithms more efficiently. In this sense, we can use the "Noisy Intermediate Scale Quantum Systems" (NISQ) **[6]** to serve the purpose. We can run the less complex quantum subroutines of a big qAI/qML in these kinds of quantum computers and use the results in the main qAI/qML problem-solving pipeline. This way we create a classical-quantum hybrid problem-solving eco-system in AI/ML space. We further can optimize the quantum subroutines at the quantum circuit level using ZX-calculus **[7]**. The optimized quantum circuits are less prone to the noisy results as the NISQ has to handle a lesser number of quantum gates calculations as compared to the original unoptimized quantum circuit. In this paper, we address an interesting problem in natural language processing (NLP) know as POS tagging **[8]** in a classical-quantum hybrid AI eco-system.

Parts of speech (POS) tagging **[8]** is a very important task in Natural Language Processing (NLP) **[9]**. POS tagging is the process of assigning one of the parts of speech to a given word. Parts of speech include nouns, verbs, adverbs, adjectives, pronouns, conjunction, and their sub-categories. For example, considering the English tag-set "Penn Treebank" at the University of Pennsylvania **[10]**:

$$word: Paper \rightarrow tag: NN$$
$$word: Famous \rightarrow tag: JJR$$
$$word: Go \rightarrow tag: VB$$
$$word: Chair \rightarrow tag: NN, VB$$

(1)

Where the POS tags *"NN"*, *"JJR"*, *"VB"* are described as *"noun"*, *"adjective/comparative"*, *"verb"* respectively. Note that some words can have more than one tag associated with it. For example, "Chair" can be "NN" or "VB" depending on the context. The POS tagger is a function that does the tagging job. Taggers use several kinds of information like dictionaries, lexicons, rules, etc. Dictionaries have a category or categories of a particular word. A word may belong to more than one category, like "Chair". Taggers use probabilistic information to solve this ambiguity. There are mainly two types of taggers: rule-based taggers **[11]** and stochastic taggers **[12]**. Rule-based taggers use hand-written rules to distinguish the tag ambiguity. Stochastic taggers are either Hidden Markov Model (HMM) **[13]** based taggers, where we choose the tag sequence which maximizes the product of word likelihood and the tag sequence probability, or cue-based taggers **[14],** where we use decision trees **[15]** or maximum entropy models **[16]** to combine probabilistic features.

Another important concept is "Tag-set". This is the set of tags from which the tagger is supposed to choose for attaching the relevant words. Every tagger is given a standard "tag-set". The "tag-set" may be coarse, such as $N\ (Noun)$, $V\ (Verb)$, $ADJ\ (Adjective)$, $ADV\ (Adverb)$, $PREP\ (Preposition)$, $CONJ\ (Conjunction)$ or fine-grained, such as $NNOM\ (Noun-Nominative)$, $NSOC\ (Noun\ Sociative)$, $VFIN\ (Verb\ Finite)$, $VNFIN\ (Verb\ Nonfinite)$ and so on. Most of the taggers use only fine-grained "tag-sets".

The runtime complexity of the proposed solution for parts of the speech tagging problem with HMM grows exponentially with the number of tags and words in the input sequence. The runtime complexity is optimized with the help of the *Viterbi* algorithm **[17]**. We further optimize the solution by formulating the problem in a quantum computing approach. In the process of building the solution, we develop a quantum version of the *Viterbi* algorithm. Additionally, we also demonstrate the optimization in implementing the solution in a quantum computer with the help of the ZX-calculus **[7]** keeping the target in the context of implementation in the Noisy Intermediate-Scale Quantum (NISQ) **[6]** systems.

## 2    PARTS OF SPEECH TAGGING WITH HMM

The HMM is a statistical Markov model **[18]** which assures to follow Markov properties. In HMM, the states are not observable, but when we visit a state, an observation is recorded that is a probabilistic function of the state. With given emission probabilities $P_{emission} = P(w_i|t_i)$ and transition probabilities $P_{transition} = P(t_i|t_{1,i-1})$, the goal of an HMM-based tagger is to maximize the following expression,

$$\max_{e \in T^W}\left\{\left(\prod P^{(e)}(w_i|t_i)\right)\left(\prod P^{(e)}(t_i|t_{1,i-1})\right)\right\} \qquad (2)$$

where $P_j^{(e)}$ is the probability of the $e^{th}$ hidden tag sequence, $w_i$ is the word at $i^{th}$ state, $t_i$ is the tag at $i^{th}$ state, $P(w_i|t_j)$ is the likelihood probability, and $P(t_i|t_{1,i-1})$ is the prior probability. With the help of emission probabilities, we generate a $T \times W$ sized emission probability matrix $B$, and a $T \times T$ sized transition probability matrix $A$. Where $T$ is the number of tags, $W$ is the number of words in a word sequence, there will be $T^W$ number of different possible tag sequences. This number grows very fast. The following **Fig.1** shows the HMM proses with a given sentence and related tags:

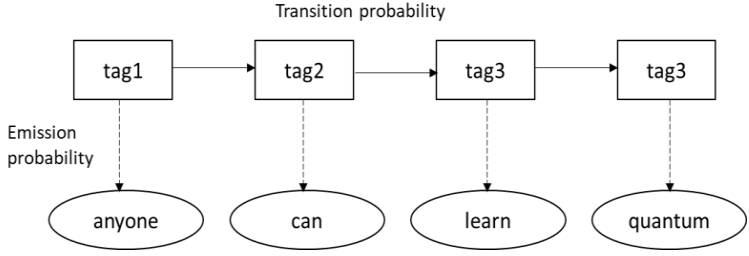

Figure 1.: HMM process with words and tags. The task is to find the most likely tag sequence for the given word sequence.

We further discuss the optimization of the runtime computational complexity of the proposed approach with the help of quantum mechanics. In the discussed approach with HMM, out of $T^W$ sequences, we have to find the sequence for which the probability is the maximum. Here, to find the maximum probability of out the $T^W$ sequences, it takes $O(T^W)$ runtime steps. We now use the following quantum version of the max finding algorithm (**Algorithm 1**), which is inspired by the algorithm of *Durr et al.* on finding the minimum value quantum mechanically **[19]**. The $quantumProbMax(V, T^W)$ takes only $O(\sqrt{T^W}) + O(W \log^2 T)$ runtime steps. Here $V$ is a list which holds all the $T^W$ probability values.

**ALGORITHM 1:** Quantum Max Probability Finding Algorithm

$quantumProbMax(V, l)$

1: initialize random element $r, 1 \leq r \leq l$
2: initialize $V$ as the vector of all classified class probabilities
3: while $\left(total\ running\ time < \left(O(\sqrt{l}) + O(\log^2 l)\right)\right)$
4:   initialize the memory as $\frac{1}{\sqrt{l}}\sum_j |j\rangle|r\rangle \to |\Phi\rangle$
5:   $quantumSearch\ (|\Phi\rangle, V, r)$
6:   if $(V[r_1] > V[r])$
7:     $r_1 \to r$
8: return $V[r], r$

Here we randomly initialize $r$, which is a uniformly chosen value from the vector $V$, where $V$ is a vector. The *while* loop terminates when the *total running time* is greater than or equal to $O(\sqrt{l})$. We then initialize the memory and call the function *quantumSearch* to apply the *Grover's Quantum Search* algorithm, which amplifies the amplitudes of all the items in $V$ whose corresponding values are greater than the threshold value $V[r]$ and mark them with the condition $(V[j] > V[r])$. Once the amplitude of these elements has been amplified, we measure on $|r\rangle$ to obtain a new threshold index $r_1$. It then returns the $r$ for the largest success probability value. The runtime of the iteration $\left(O(\sqrt{l}) + O(\log^2 l)\right)$ can be demarcated as $(22.5\ \sqrt{l} + 1.4 \log^2 l)$ **[20]**.

Additionally, there is a popular dynamic programming algorithm known as the *"Viterbi"* algorithm that provides an optimal way to find the most likely sequence of hidden states, which generates a sequence of observed events. The runtime complexity of the Viterbi algorithm is $O(T^2 W)$. In the next section, we further optimize the runtime complexity of this algorithm with the help of quantum postulates.

# 3 VITERBI ALGORITHM OPTIMIZATION IN QUANTUM APPROACH

*Viterbi* algorithm [17] is a popular algorithm for the application of dynamic programming algorithms to maximization problems which involves probabilities. The following **Algorithm 2** demonstrates the Viterbi algorithm in the classical domain [21]:

---
**ALGORITHM 2:** Classical Viterbi Algorithm
---

$classicalViterbi(O, S, \Pi, Y, A, B)$

1:   for state $i = 1,2, \dots, K$:

2:   $\quad \pi_i \cdot B_{iy_1} \to \varphi_1[i, 1]$

3:   $\quad 0 \to \varphi_1[i, 1]$

4:   for observation $j = 2,3, \dots, \varphi$:

5:   $\quad$ for state $i = 1,2, \dots, K$:

6:   $\quad\quad \max_k \left( \varphi_1[k, j-1] \cdot A_{ki} \cdot B_{iy_j} \right) \to \varphi_1[i, j]$

7:   $\quad\quad \arg\max_k \left( \varphi_1[k, j-1] \cdot A_{ki} \cdot B_{iy_j} \right) \to \varphi_2[i, j]$

8:   $\arg\max_k (\varphi_1[k, \varphi]) \to z_\varphi$

9:   $s_{z_\varphi} \to x_\varphi$

10:  for $j = \varphi, \varphi - 1, \dots, 2$:

11:  $\quad \varphi_2[z_j, j] \to z_{j-1}$

12:  $\quad s_{z_{j-1}} \to x_{j-1}$

13:  $x_\varphi \to X$

14:  return $X$

The above **Algorithm 2** generate the most likely hidden states $X = (x_1, x_2, \dots, x_\varphi)$; $x_n \in S = \{s_1, s_2, \dots, s_K\}$, where $S$ is the set of sequences of states. $Y = (y_1, y_2, \dots, y_\varphi)$ is the set of observations generated with $y_n \in O = \{o_1, o_2, \dots, o_N\}$, where the observation space $O$ possesses $N$ number of possible observations. $\Pi = (\pi_1, \pi_2, \dots, \pi_K)$ is the set of initial probabilities, where $\pi_i$ stores the probability that $x_i = s_i$. $A$ is $K \times K$ transition matrix, where $A_{ij}$ stores the transition probability of the transition from state $s_i$ to state $s_j$. $B$ represents a $K \times N$ emission matrix, where $B_{ij}$ stores the probability of observing $o_j$ from $s_i$. Each element $\varphi_1[i, j]$ of $\varphi_1$ stores the probability of the most likely path so far $X = (x_1, x_2, \dots, x_j)$ with $x_j = s_j$ that generates $Y = (y_1, y_2, \dots, y_j)$, and the element $\varphi_2[i, j]$ of $\varphi_2$ stores $x_{j-1}$ of the most likely path so far $X =$

$(x_1, x_2, \ldots, x_{j-1}, x_j = s_i) \, \forall j, 2 \leq j \leq \varphi$. We then fill the table entries of $\varphi_1[i,j], \varphi_2[i,j]$ with the increasing the order of $K \cdot j + i$ as,

$$\max_k \left( \varphi_1[k, j-1] \cdot A_{ki} \cdot B_{iy_j} \right) \to \varphi_1[i,j], \tag{3}$$

$$\arg \max_k \left( \varphi_1[k, j-1] \cdot A_{ki} \cdot B_{iy_j} \right) \to \varphi_2[i,j]. \tag{4}$$

The below **Algorithm 3** is the optimized version of the *Viterbi* algorithm in a quantum computing approach. As compared to the maximum finding function in the classical *Viterbi* algorithm, we use the *quantumMax* function to find the max value and the associated index quantum mechanically.

**ALGORITHM 3:** Quantum Viterbi Algorithm

$quantumViterbi(O, S, \Pi, Y, A, B)$

1:    for state $i = 1, 2, \ldots, K$:

2:        $\pi_i \cdot B_{iy_1} \to \varphi_1[i, 1]$

3:        $0 \to \varphi_1[i, 1]$

4:    for observation $j = 2, 3, \ldots, \varphi$:

5:        for state $i = 1, 2, \ldots, K$:

6:            $quantumMax_k \left( \varphi_1[k, j-1] \cdot A_{ki} \cdot B_{iy_j}, K \right)[0] \to \varphi_1[i,j]$

7:            $quantumMax_k \left( \varphi_1[k, j-1] \cdot A_{ki} \cdot B_{iy_j}, K \right)[1] \to \varphi_2[i,j]$

8:    $quantumMax_k (\varphi_1[k, \varphi], K)[1] \to z_\varphi$

9:    $s_{z_\varphi} \to x_\varphi$

10:   for $j = \varphi, \varphi - 1, \ldots, 2$:

11:       $\varphi_2[z_j, j] \to z_{j-1}$

12:       $s_{z_{j-1}} \to x_{j-1}$

13:   $x_\varphi \to X$

14:   return $X$

The process in **Algorithm 3** is similar to the **Algorithm 2** except, we fill the table entries of $\varphi_1[i,j], \varphi_2[i,j]$ with the increasing the order of $K \cdot j + i$ as,

$$\underset{k}{quantumMax} \left(\varphi_1[k, j-1] \cdot A_{ki} \cdot B_{iy_j}, K\right)[0] \to \varphi_1[i,j], \quad (5)$$

$$\underset{k}{quantumMax}\left(\varphi_1[k, j-1] \cdot A_{ki}, K\right)[1] \to \varphi_2[i,j]. \quad (6)$$

In this quantum version of the *Viterbi* algorithm, we use the quantum max finding function (**Algorithm 1**), which gives quadratic runtime performance improvements at $T(W-1)$ places during the max finding at these cells in the trellis diagram. For each hidden tag cell in the trellis diagram, there will be $T$ probability values, and there are $T(W-1)$ such cells ($T$ cells will have only one probability value). At each hidden tag cell in the trellis diagram, the quantum version of max finding **Algorithm 1** takes only $\left(O(\sqrt{T}) + O(log^2 T)\right)$ as compared to the classical $\left(O(T) + O(log^2 T)\right)$ runtime steps to find the maximum probability from $T$ probability values. Therefore, for $T(W-1)$ such hidden tag cells, it will take only $\left(O(T^{1.5}(W-1)) + O(T(W-1)log^2 T)\right)$ runtime steps as compared to the classical $\left(O(T^2(W-1)) + O(T(W-1)log^2 T)\right)$ runtime steps.

The proposed approach of parts of tagging with the quantum realm demonstrates significant runtime performance improvements over the classical approach. Now, the next target is to optimize the approach from an implementation point of view in a real quantum computer. At present, NISQ **[6]** seems to be a very promising system to implement such quantum-based NLP algorithms. Our quantum approach uses Grover's search at the heart of the framework, so by optimizing Grover's search quantum circuit we can make the proposed approach more optimal implementable for NISQ. In Section 4, we discuss the quantum circuit optimization with ZX-Calculus **[7]** for our proposed quantum approach of parts of speech tagging.

4   NUMBER OF GATES OPTIMIZATION WITH ZX-CALCULUS

The ZX-calculus helps in optimizing a quantum circuit by reasoning about the linear mapping between qubits using a graphical language model representation **[22]**. The ZX-diagrams characterize the linear mapping between qubits, analogously the way quantum circuits represent the unitary mapping between qubits in the quantum circuits. *Coecke et al.* have introduced the "ZX-calculus" in 2008 **[7]**. They also introduced the associated term "spider", which is the fundamental concept in the ZX-calculus for rewriting the rules and is a strong complementary. The ZX-calculus is universal, so we can represent any linear mapping between qubits as a ZX-diagram. A ZX-calculus comprise of spiders that are connected by flexible wires (wires can cross and curve). The spiders can be green and red colored nodes in the ZX-diagram. The color green represents the computational basis and the color red represents the Hadamard-transformed basis. Apart from this, the yellow square node represents the Hadamard node, which at all times connects to exactly two wires.

The ZX-calculus comes with a certain set of rewrite rules, which are used to perform the calculations in the graphical language itself. The building blocks of the ZX-calculus are known as "generators", which are some specific representations of states, linear isometries, unitary operations, the Hadamard-transformed basis, and the computational basis ($|0\rangle, |1\rangle$) based projections. The generators can be composed either sequentially or in parallel. Some of the important rewrite rules are like, "same color adjacent spiders merge", "arity-2 spiders are identical", and "Hadamard changes the color of the spiders". Also, the topology has no meaning in the context, so if the two diagrams consist of the same generators connected in the same way, the two diagrams represent the same linear operation context. The ZX-diagram of the following GHZ (Greenberger–Horne–Zeilinger) state **[22]** quantum circuit demonstrated as below in **Fig.2**:

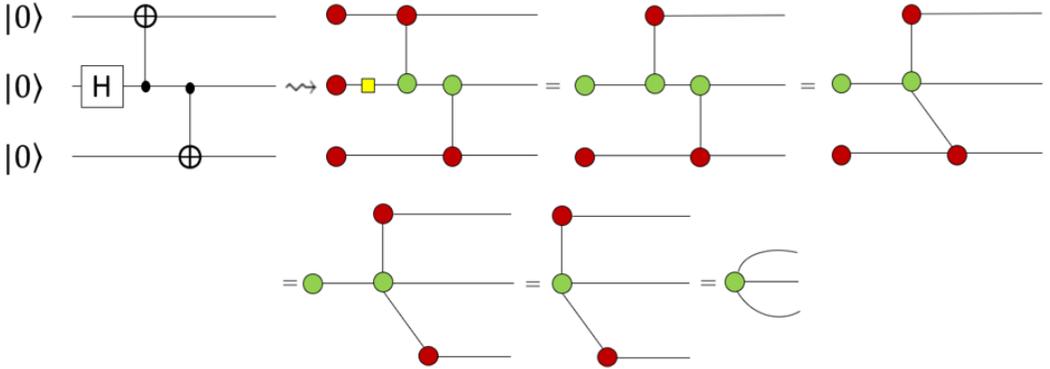

Figure 2. Demonstration of optimization of the GHZ-state quantum circuit with ZX-calculus.

GHZ-state is a certain type of entangled quantum state in the area of quantum information theory, which was first studied by Daniel Greenberger, Michael Horne, and Anton Zeilinger in 1989 **[23]**. These rules rewriting process can be automatically done by using the tool "Quantomatic" **[24]**. In our proposed quantum way of parts of speech tagging, the *quantumMax* function uses Grover's search, so we optimize its quantum circuit using the ZX-calculus. The formulation of the quantum version of *Viterbi* algorithm uses *quantumMax* function $T(W-1)$ times. With every call to *quantumMax* function, it executes Grover's quantum search algorithm. Therefore, there are $T(W-1)$ calls to Grover's search algorithm. In **[25]**, the researchers have shown that the quantum circuit of Grover's search with 5 qubits inputs has 336 quantum T-gates (T-gate is equivalent to Rz-gate for the angle $\frac{\pi}{4}$) in total. Using the ZX-calculus, the number of T-gates in the quantum circuit is optimized to 166 quantum T-gates. In the context of our quantum version of *Viterbi* algorithm, with a specific case, when the number of tags $T=5$ and number of words $W=5$, there will be $336(T(W-1)) = 316 \times (5(5-1)) = 6320$ T-gates for execution in a quantum computer. After applying the ZX-calculus, it is optimized to $166(T(W-1)) = 166 \times (5(5-1)) = 3320$ T-gates in total to execute on the quantum computer. This results in ~ 47.47% reductions in the total number of T-gates execution in a quantum computer.

## 5  CONCLUSION

In this paper, we have demonstrated the optimization of the parts of speech tagging formulation in a quantum approach. During the process of optimization, we formulated a quantum version of the *Viterbi* algorithm. Our quantum proposed approach shows quadratic runtime performance improvements as compared to the classical counterpart. We also demonstrate that with the help of ZX-calculus, our quantum approach can be implementable with 47.47% less number of quantum T-gates in NISQ. In this research, the analysis and results make us more close to harnessing the NISQ system from an NLP implementation point of view in noisy quantum systems (NISQ) until we encounter with a fully functional commercial quantum system. Our research also demonstrates and classical-quantum hybrid AI-eco system, which enables us to motivate for harnessing the present-day quantum systems along with the powerful classical systems to solve complex AI problems.